\documentclass[12pt]{article}
\usepackage{amsmath}
\usepackage{amssymb}
\usepackage{epsfig}
\usepackage{cite}
\usepackage{color,colordvi}
\newcommand{\eq}{\begin{equation}}
\newcommand{\eqx}{\end{equation}}
\newcommand{\eqn}{\begin{eqnarray}}
\newcommand{\bi}{\begin{itemize}}
\newcommand{\eqnx}{\end{eqnarray}}
\newcommand{\ei}{\end{itemize}}

\newcounter{hran}

\def\MSbar{\relax\ifmmode\overline{\rm MS}\else{$\overline{\rm MS}${ }}\fi}

\begin{document}

\begin{center}

\hfill CERN-PH-TH/2010-082\\
\vskip -.1 cm
\hfill LMU-ASC 26/10
\vskip -.1 cm
\hfill MPP-2010-49\\
\vskip -.1 cm
\hfill IFT UAM/CSIC 10-37\\

{\Large\bf Self-Completeness of Einstein Gravity.}

\vspace{0.5cm}

\end{center}

\begin{center}

{\bf Gia Dvali}$^{a,b,d,c}$ and {\bf Cesar Gomez}$^{e}$

\vspace{.6truecm}

\vspace{.2truecm}

{\em $^a$Arnold Sommerfeld Center for Theoretical Physics\\
Department f\"ur Physik, Ludwig-Maximilians-Universit\"at M\"unchen\\
Theresienstr.~37, 80333 M\"unchen, Germany}


{\em $^b$Max-Planck-Institut f\"ur Physik\\
F\"ohringer Ring 6, 80805 M\"unchen, Germany}


{\em $^c$CERN,
Theory Division\\
1211 Geneva 23, Switzerland}


{\em $^d$CCPP,
Department of Physics, New York University\\
4 Washington Place, New York, NY 10003, USA}

\vspace{.2truecm}

{\em $^e$
Instituto de F\'{\i}sica Te\'orica UAM-CSIC, C-XVI \\
Universidad Aut\'onoma de Madrid,
Cantoblanco, 28049 Madrid, Spain}\\

\end{center}


\centerline{\bf Abstract}
 
 We argue, that in Einsteinian gravity the Planck length is the shortest length of 
 nature, and any attempt of resolving trans-Planckian physics bounces back to macroscopic distances due to black hole formation. 
  In Einstein gravity trans-Planckian propagating quantum degrees of freedom cannot exist, instead  they are equivalent to the classical black holes that are fully described by lighter infra-red degrees of freedom and give exponentially-soft contribution into the virtual processes.   
 Based on this property we argue that pure-Einstein (super)gravity and its high-dimensional generalizations are  self-complete in deep-UV, but not in standard Wilsonian sense.  
  We suggest that 
certain strong-coupling limit of string theory is  built-in in pure Einstein gravity, whereas the role of weakly-coupled string theory limit is to consistently couple gravity to other particle species, with their number  being set by the inverse string coupling.    
We also discuss some speculative ideas generalizing the notion of non-Wilsonian 
self-completeness  to other theories, such as the standard model without the Higgs.

\noindent

\vskip .4in
\noindent


\section{The  Shortest  Scale of Nature}

   In gravity  the Planck length  is {\it the shortest}  length-scale of nature, and any attempt of probing the shorter distances will instead probe the larger length-scales. 
  
   We start with the pure Einsteinian theory of gravity in four-dimensions, in which below the 
   Planck scale the only propagating degree of freedom is a massless spin-2 graviton, 
   $h_{\mu\nu}$. 
   No other propagating species are assumed at this point. All the sources, other than the ones composed
   out of gravitons, will be considered as external sources that are not associated with any
   new degrees of freedom.  This requirement uniquely fixes the action in form of the Einstein-Hilbert action,  
   \begin{equation}
 S_{EH} \, = \,    \int \, d^4x \, M_P^2 \, \sqrt {-g} \, {\mathcal R}\, .
     \label{EH}
     \end{equation}
   For definiteness, we have set the cosmological constant to zero.  In this way, we shall always 
   consider observers on asymptotically-flat  spaces.  
   
     We shall first clarify the field-theoretic meaning of the Planck mass $M_P \, \sim \, 10^{19}$GeV and of the 
corresponding Planck length,  $L_p\, \equiv \, M_P^{-1} \, \sim 10^{-33}$ cm.   The overall numerical factors of order one will be ignored throughout the paper.

        In pure Einstein gravity, the Planck scale plays the central role. 
     It defines the coupling of graviton to energy momentum sources {\it universally}, 
  \begin{equation}
   h_{\mu\nu} \, {T^{\mu\nu}  \over  M_P}  \, ,   
  \label{gravitoncoupling}
  \end{equation}
  where $T_{\mu\nu}$ is an arbitrary energy-momentum tensor.  The important fact is, that the 
  above universality property is true to all orders in non-linear interactions of graviton. 
  That is, one can either think of the above coupling as the coupling to an external source, or as  
a self-coupling of graviton to its own energy-momentum tensor in a given non-linear order. 
   Due to the above crucial property, Einstein's gravity viewed as a quantum field theory possesses an universal strong coupling scale, $M_P$.  
   This fact shall play the key role in what follows. 
  
    The existing common knowledge about Einstein gravity is, that it becomes unapplicable in deep-UV, at distances  $L \ll \, L_P$ and must be completed by a more powerful theory that will restore consistency at sub-Planckian distances. 
   
    We wish to question the above statement and suggest, 
    that the  pure Einstein's gravity is {\it  self-complete}  in deep-UV. In other words, we argue that for restoring consistency 
    no new propagating degrees of freedom are necessary at energies $\gg \, M_P$, and moreover,  
 even if one tries to introduce such new states, they will not have any physical meaning, since the corresponding distances can never be probed.  All the information that such new states can in principle  carry, will be identical to the information carried by the semi-classical macroscopic  black holes of the same mass, whose properties are completely determined by the  IR gravity.

  The reason behind our claim is,  that in Einstein's gravity, $L_P$ represents the absolute lower
  bound on any distance that can ever be resolved. Distances  $L \, \ll \, L_P$, cannot be probed, 
  {\it in principle}.
   
  A version of the above statement  sometimes goes under the name of Generalized Uncertainty Principle \cite{GUP1, GUP2}.   
  
    More precisely, any attempt of resolving physics at the distance scales $L \,  \ll \, L_p$, will inevitably bounce us back to much larger distances $L_P^2/L   \,  \gg \, L_P$, which are completely insensitive to any short-distance physics and are entirely governed by the massless graviton, which is the only king  at any scale longer than the Planck length.  
   
    Namely,   physics that one can decode at  sub-Planckian distances is identical to the physics at macroscopic distances, 
  \begin{equation}
   L \, \leftarrow \rightarrow   \,  {L_P^2 \over L} \, .
   \label{central}
  \end{equation}   
  
    The fundamental reason for such an obstacle is the existence  of black holes (BH). 
  BH-formation interferes with any attempt of extracting information from beyond the Planck length and produces 
  an insuperable barrier.  In fact, harder we try to go beyond  $L_P$,  with a larger and more classical BH we shall end up.  Physics of such a BH has nothing to do with short distances and is entirely governed by the infra-red (IR) gravity.   
   
  In other  words,  the key reason for our claim can be formulated  in the following way,
  \begin{equation}
  {\rm Deep-UV~Gravity  ~ =  ~ Deep-IR ~Gravity } \,. 
  \label{central}
  \end{equation}
  Some crucial aspects of the above connection has been stressed by the previous authors \cite{Banks, Susskind}. 
  However, we shall suggest that it is absolute in field-theoretical sense.  Namely, in Einstein gravity trans-Planckian propagating quantum degrees of freedom cannot exist, instead they are mapped on (non-propagating) classical states, fully described by the dynamics of  lighter propagating IR degrees of freedom, such as the massless graviton.  Any attempt of integrating-in trans-Planckian quantum fields  that avoid such correspondence is bound to fail. 
Since,  dynamics of quantum field theories is formulated in terms of propagating degrees of freedom,   our conclusion is, that Einstein's gravity is self-UV-complete, but in the sense that is 
different from the notion of standard Wilsonian UV-completeness.  
 
   In order to see how the concept of the minimal length arises, consider a generic thought experiment that attempts to resolve the physics at distance $L$.  An elementary act of such a measurement is a scattering process in which one has to localize the minimum amount of energy $E \, > \, 1/L$ within the space-time 
   box of size $L$.  The corresponding  Schwarzschild radius of such a localized energy 
  portion is, 
  \begin{equation}
  R(L) \, = \, L_P^2/L \, .
  \label{radius}
  \end{equation}
  Notice, that for $L \, \ll \, L_P$ the above Schwarzschild radius exceeds both $L$ and the Planck length. 
  Thus, any attempt of probing length scales $L \, \ll \, L_P$ will require localization of energy 
  within the radius much smaller than the corresponding Schwarzschild, 
  $R (L) \, \gg \, L_P$.  The  corresponding act of measurement thus will lead to a formation of a macroscopic  {\it classical}  BH, way before it has any chance of probing distance $L$.
  
    The above conclusion is completely  insensitive to what formally happens to the gravitational dynamics in the trans-Planckian region $L \, << \, L_P$. 
The BH shield is turned on way before this dynamics has any chance to get excited. 

   To put it in different terms, the maximal  information that can be extracted from the measurements of a sub-Planckian distance $L$  is equal to the information that can be  encoded at the horizon of a classical BH of size $L_P^2/L$.  
   
   The above reasoning is in full accordance with the ideas of holography\cite{holog}.  
 It can only be violated if the theory possesses 
an agent that could violate energy positivity condition.  Then, using such an agent,  one could encode (and extract) information at arbitrarily short distances, without paying the energy price.

   Let us suppose we attempt to change the laws of gravity dramatically at distances 
   $L \, \ll \, L_P$.   For this we have to introduce new gravitational degrees of freedom with masses 
   $m \, = \ L^{-1}\, \gg \, M_P$.  In the other words, we try to introduce new poles in the graviton propagator 
   at some $p^2 \, = \, L^{-2} \,  \gg \,  M_P^2$.  Naively,  such poles will change gravitational dynamics at distances $L$, but this is a complete illusion, since corresponding change can never be probed. 
   Any observer that will attempt to probe the physics of the trans-Planckian pole, will not learn anything new other than what he/she can learn in Einsteinian gravity at distance 
   $L_P^2/L$. 
   
    We shall now attempt to give a field theoretic perspective of the above phenomenon.
   A related work  with applications to particular UV-completions  of gravity shall  appear in\cite{sarah}.   
       
     We shall discuss this on a concrete example.  Let us say, we modify the graviton propagator 
   by adding a new trans-Planckian pole  at $1/L$,     
     \begin{equation}
    {1 \over M_P^2} \, T^{\mu\nu} \, \langle h_{\mu\nu}h_{\alpha\beta} \rangle \,   t^{\alpha\beta}\, =  \, 
   {1 \over M_P^2} \, \left ( {T_{\mu\nu}t^{\mu\nu} \, - \, {1 \over 2} 
   T_{\mu}^{\mu} t_{\nu}^{\nu}
     \over p^2 } \, 
      + \, {a T_{\mu\nu}t^{\mu\nu} \, - b\, {1 \over 3} 
   T_{\mu}^{\mu} t_{\nu}^{\nu}
     \over p^2 \, + \, (1/L)^2}\,  \right ) \, ,  
     \label{spectralrep}
   \end{equation}
   where we have convoluted the propagator with the two sources,  $T_{\mu\nu}$ and $t_{\mu\nu}$.   The pole at $p^2 \, = \, 0$ in the first term corresponds to the Einsteinian massless graviton.  
  The parameters $a$ and $b$ are fixed according to the spin of the new pole
   ($a=b$ for spin-2,  and $a=0, b< 0$ for spin-0).  
 
    In order to probe a pole at $p^2 \, = \, L^{-2}\, \gg \, M_P^2$, we need to consider an experiment with the momentum-transfer $\sim \, L^{-1}$. In any such process
    we need to localize energy $1/L$ within the distance $L\, \ll \, L_P$.   But, this is impossible 
  without first forming a classical BH of size  $R(L) \, \gg \, L_P$. 
    For example, we can scatter  gravitons with an impact parameter $L$ and center of mass energy $1/L$. 
    But since the Schwarzschild radius is much larger than the impact parameter, 
    \begin{equation}
  R(L) \, = \, L_P^2/L \, \gg \, L_{impact} \, = \, L \, , 
  \label{impact}
  \end{equation}    
    the classical 
    BH will form way before the scattering gravitons have any chance of approaching the distance  $L$.  Thus,  a trans-Planckian pole in the  graviton propagator remains completely shielded by  the BH barrier and is unaccessible, in principle. 
    Or to be more precise, accessing such a pole is the same as accessing the classical BH of the same mass, and thus former cannot carry any other information that the latter. 
    
    Thus,  our attempt to integrate in a propagating quantum degree of freedom of trans-Planckian 
mass failed and we ended up with a classical BH instead.  This means, that representation of the graviton propagator in the form  (\ref{spectralrep}) for  trans-Planckian poles is  inconsistent.  Contribution from such poles must be exponentially suppressed. 
 We can estimate  the required suppression factor (up to a numerical coefficient in the exponent) as,    
 \begin{equation}
 {\rm e}^{-(L_P/L)^2} \, .
 \label{expfactor}
 \end{equation}
 The necessity of this factor can be understood  at least in two ways.  First, it can be interpreted as the entropy suppression, e$^{-S}$, where  $S \, = \, (L_P/L)^2$ is the Bekenstein-Hawking  entropy. 
 Secondly, it can be interpreted as the Boltzmann suppression in the evaporation of a classical BH.  
 
 In order to explain the latter suppression, let us denote the (would-be) degree of freedom corresponding to the massive pole,  $p^2 \, = \, L^{-2}$,  by $\phi_{\mu\nu}$. The spin ($2$ versus $0$) is unimportant for 
 the present discussion and we shall leave it unspecified.   Let  the source $T_{\mu\nu}$,  to which 
 $\phi$ is coupled, be an energy momentum tensor of some light (not necessarily massless ) particle $q$.   Then, the interaction vertex $\phi_{\mu\nu}T^{\mu\nu}$   
sets the decay rate of $\phi$ into a particle anti-particle pair, 
\begin{equation}
\phi \, \rightarrow \, q \, + \, \bar{q} \, .
\label{phidecay}
\end{equation}
 As long as $L^{-1} \, \ll \, M_P$, this is an ordinary quantum decay of a heavy particle into the lighter ones. However, for 
 $L^{-1} \, \gg \, M_P$, the same vertex describes evaporation of a classical BH of mass 
 $1/L$ into a single particle-anti-particle pair.  This process is obviously suppressed by the 
 Boltzmann factor e$^{-(1/LT)}$, since the particle pair has energy $1/L$, which exceeds the 
 Hawking temperature
 $T \, = \, L/L_P^2$.   This Boltzmann suppression matches (\ref{expfactor}).    
 
 Thus, the two conclusions follow from the above analysis. First, the effects of the trans-Planckian poles are exponentially softened, and secondly, such poles no longer describe quantum propagating degrees of freedom, but rather the classical states.      
    
      The above reincarnation of  trans-Planckian degrees of freedom  into the classical states is one of the key points of our analysis.  What we are observing is, that in gravity  there are no propagating quantum degrees of freedom above the Planck mass, instead they become classical states that are fully described by other lighter propagating degrees of freedom.              
      
      Generic quantum field theories are defined by quantum propagating degrees of freedom and by their interactions.  Field theories also describe classical states, such as solitons or other classical solutions.  The defining property of the classical states is, that they are not  independent entities, and at least in principle can be fully described by quantum degrees of freedom.   For example, solitons can be understood as  coherent superpositions of quantum degrees of freedom with large occupation numbers.  
     
      The intrinsic property  of classical states is that they cannot probe distances shorter than the characteristic wavelength of their constituent quantum particles, which is typically given by the size of the classical configuration in question.  For example, the size of  a 't Hooft-Polyakov magnetic monopole is given by the Compton wave-length of the massive gauge fields.  Because of this, 
despite the fact that monopole can be much heavier than the gauge field, the former is no better probe of the short-distance physics than the latter.     

  The above statements are true both for gravity and for other field theories. 
  However,  the crucial peculiarity of Einstein gravity is  the following. In ordinary quantum      
 field theories, with suitable arrangements, one can include arbitrarily heavy quantum degrees of freedom.  By making their mass higher, heavy degrees of freedom probe shorter and shorter distances.  Of course, they do gradually decouple from the low energy processes, but their effects can in principle 
 be detected in precise measurements at arbitrarily large  distances.    
  
   The story in Einstein gravity is dramatically different. By becoming heavier than $M_P$ particles  simply stop existence as independent quantum {\it degrees of freedom} and become classical {\it states}. 
  These classical states are no longer independent entities, but instead are fully described by 
  other already-existing light fundamental degrees of freedom, such as the massless graviton.  
    
     This transition of the heavy would-be degrees of freedom into non-fundamental classical 
   states  is intrinsic property of gravity, and the key to its self-completeness \footnote{As said above, the Einsteinian reincarnation of the deep UV poles  into classical states automatically softens trans-Planckian effects,  which become exponentially-suppressed. This softness which is one of the main symptoms of self completeness is reminiscent of the softness of trans-Planckian string theory amplitudes.}. 
    From this behavior it also follows that at the boundary of the two regimes,  some quantum degrees of freedom  of mass $M_P$ must be present in Einstein gravity.  
    
     Thus,  the built-in spectrum of quantum degrees of freedom  of Einstein gravity includes massless graviton plus new quantum degrees of freedom in a narrow mass interval around $M_P$.  As we shall show,  existence of the latter  degrees of freedom  is not an additional assumption, but follows from the smooth transition between the quantum particles and classical states.  However, their presence plays essentially no role neither in deep-UV nor in deep-IR.  
 The rest of the states in Einstein gravity are not fundamental and are described by the dynamics of the massless graviton.     
 
  This findings lead us to the conclusion that Einstein gravity is self-complete in a sense that 
  is very different from the standard notion of  the  Wilsonian completeness.  
 
  In Wilsonian sense a quantum field theory is defined as a relevant perturbation of an  UV fixed point CFT. The UV CFT sets the real degrees of freedom of the theory as well as those unitarizing the high energy scattering S matrix. In this frame, given a particular low energy physics,  the corresponding UV completion is obtained by embedding that IR dynamics into a quantum field theory flowing in the UV to a CFT fixed point. As already pointed out the essential aspect of gravity is the existence of a BH barrier for resolving scales smaller than Planck length. A natural consequence of this barrier is to unitarize high energy scattering amplitudes using black hole production. This led to the hypothesis known as asymptotic darkness \cite{Banks} and to postulate BHs as the real UV states of the theory. However this potential UV description of quantum gravity in terms of BHs does not easily fit  with any UV Wilsonian CFT. Indeed the essential property of BHs is to carry entropy and therefore any UV CFT Wilsonian description of quantum gravity - consistent with the BH barrier to short scales resolution - should be able to account for the BH Bekenstein entropy. The BH entropy for asymptotically flat BHs in generic dimension $d$ scales with energy as $E^{\frac{d-2}{d-3}}$, while the entropy for a CFT scales as  $E^{\frac{d-1}{d}}$. This mismatch for the entropy formula for asymptotically flat BHs indicates that quantum gravity is not a Wilsonian quantum field theory. The situation changes drastically in the case of negative cosmological constant. For asymptotically AdS black holes the Bekenstein entropy goes like $E^{\frac{d-2}{d-1}}$ and therefore we can account for that entropy using a CFT in one less dimension. This leads to the famous AdS/CFT \cite{ADS} definition of quantum gravity in five dimensions in terms of the $N=4$ SYM CFT.

 Independently of what could the microscopic theory that accounts for the BH entropy be, the crucial consequence of describing the UV quantum gravity in terms of BHs of masses bigger than $M_P$ is that those UV degrees of freedom are - in contrast to what happens in Wilsonian-complete  quantum field theories  - perfectly well defined low energy states of the theory. Therefore we can map the UV degrees of freedom into the space of states describing gravity in the IR. It is in this sense that Einstein gravity is self-complete, although not Wilsonian.

   Although the  UV/IR transformation $L \rightarrow \frac{L^2_P}{L}$ bounces deep UV probes into classical macroscopic BHs,  we may wonder how this correspondence works near the Planck length itself. Fortunately, thanks to the fact that in pure Einstenian gravity BHs evaporate, we can define the relevant quantum degrees of freedom at the Planck length by parametrically reducing the BH mass until reaching a quantum mechanical regime with the corresponding Compton length bigger than the Schwarzschild radius. This happens when the BH becomes of mass $M_P$. 
  As we shall explain in more details later,  this crossover between the two regimes implies   existence  of propagating quantum degrees of freedom with mass around $M_P$.  
 Thus,  we can conclude that in addition to macroscopic BHs the UV description of quantum gravity requires inclusion of quantum species of mass $M_P$.   We shall discuss later the possible role of 
 these quantum species in connection to string theory and to microscopic description of BH entropy.

In summary although  the concepts of $L_P$ being a minimal length \cite{GUP2} as well as UV-IR connection through the BHs \cite{Banks} \cite{Susskind} have been  around for some time,  the goal of the present paper is to push these concepts to certain extreme and to show that they take us to 
self-completeness of Einstein gravity.
We do this by deriving the above concepts from the quantum field theoretic perspective, basing our reasoning  on fundamental notions
such as local propagating degrees of freedom and  the scattering amplitudes. This language allows us to circumvent secondary (but otherwise very important) issues,
e.g., such as BH information  loss, and to unambiguously  identify the true physical meaning of trans-Planckian degrees of freedom as of classical IR states, which is
the key for understanding the self-UV completeness of gravity.   We see that any propagating quantum degree of freedom when being pushed into the trans-Planckian region becomes a non-propagating classical state belonging to the deep IR sector of the theory.  In this way, there are no 
poles on the complex plane that are able to probe short distances.  In what follows we shall  discuss  this from various angles.

  \section{Being Patient:  Can Trans-Planckian Physics be Probed by Waiting Longer?}
    
      As we have seen from the previous section,  BHs make it impossible to probe 
  distances smaller than the Planck length in any measurement process.  The question we would like to ask now is,  whether  it is possible to circumvent the BH barrier by waiting a long enough time. 
 To formulate the question more precisely,  can we probe a new  heavy pole   $p^2 \, \gg \, M_P^2$, by  waiting for the final stages of the BH evaporation? 
 
     In general, the physics of the heavy particles,  of mass  $M_X$, can be probed in the following two ways:
     
  1) Observe the processes among the light fields mediated by the high-dimensional operators generated after integrating-out the  heavy quanta; 
  
  or 
  
  2) Detect a direct production of heavy states in the high-energy processes. 
 
   As we shall now see, none of the above is possible for $M_X \, >> \, M_P^2$. 
 We consider the two options separately. 
 
 \subsection{Can Operators Induced by Trans-Plankian  Quanta Serve as Probes of 
 Trans-Planckian Physics? }
 
  In many cases  the processes mediated by  high-dimensional  operators induced by heavy states can be probes of high-energy physics.  The well known example is the proton decay mediated by the 
 baryon and lepton number violating operators, such as, 
 \begin{equation}
                                          {qqql \over M_X^2} \, , 
 \label{protondecay}
   \end{equation}
 where $q$ and $l$ stand for quark and lepton fields respectively.   For example, in standard grand unified theories (GUTs), such operators are generated by the exchange of $X$ and $Y$ gauge bosons and colored Higgs states of mass $M_X \, \sim \, 10^{16}$GeV.  Despite of the huge 
suppression, the operators of the above sort represent direct low-energy probes of the GUT-scale physics, since the tiny decay rate  can be overcompensated by the huge number of baryons in the sample and by the possibility of performing observations over  long time-scales. 

 Can a similar reasoning be applied to the trans-Planckian physics, at least in principle? 
The answer to this question is negative.  As we have explained, any quantum of  mass  $M_X\, \gg \, M_P$ is no longer a perturbative state, but rather is a macroscopic object, a {\it classical} BH.   This fact immediately implies the following.  

First, in accordance with (\ref{expfactor}), the operators obtained by integrating out such a state must be  exponentially suppressed at least by the entropy (or Boltzmann) factor $e^{-S}$, where $S \, = \, (M_X/M_P)^2$ is the Bekenstein-Hawking entropy.  
 For example, the operator (\ref{protondecay}) can be generated as a result of collapsing 
 two quarks into a virtual classical BH with the subsequent evaporation of the latter into 
 a quark and a lepton.  As we have discussed  earlier in the paper, the effective form-factor describing evaporation of a classical BH into  a two-particle final state, must be suppressed by the 
 Boltzmann factor, e$^{-M_X/T}$, which gives  (\ref{expfactor}).

More importantly, since it is a classical BH, by BH no-hair theorem \cite{nohair}, it cannot be distinguished from any other BH of the same mass and the spin, obtained by collapse of  the low energy particles.  Therefore, an operator obtained by integrating out such a trans-Planckian state cannot be distinguished from the analogous operator obtained by integrating out any other classical BH of the same characteristics.  Since the latter object obviously cannot probe any trans-Planckian physics, the same applies to the  former one. 
In conclusion, operators mediated by trans-Planckian quanta, are unable to give any information about the deep-UV physics, in principle.

\subsection{Direct Production of trans-Planckian Quanta in BH-Evaporation?}

 Another question is, can one probe trans-Planckian degrees of freedom by their direct production in BH evaporation? 
  Again,  the answer to this question is negative.  The reason is simple.  First,   by conservation of energy,  BH can produce a particle of mass $M_X$ only until its mass  drops below $M_X$.
 But, because $M_X \, \gg \, M_P$, the  BH of corresponding mass is a classical  BH of temperature   $T_H \, = \, M_P^2/M_X\,  \ll \, M_X$.  Thus, the trans-Planckian state of the 
 mass $M_X$ itself is a classical BH of the same mass and vice-versa.  This closes the issue.  
 Since, first the production of such a heavy state will be suppressed at least by a Boltzmann factor  $e^{-(M_X/M_P)^2}$. 
 And  secondly,  since the state itself is a classical BH of the same mass, it will not carry any message about  the deep-UV physics, but rather only about the  IR physics corresponding to distances $M_X/M_P^2$. 
 
 \subsection{Jumping into a  Black Hole?}
 
  Finally, we briefly note,  that jumping into the BH and trying to probe physics of singularity 
  will not give any new information about the trans-Planckian physics.  An observer falling 
  towards the singularity is not in any respect in a better position to perform the measurement experiment 
  than a flat space observer.  If he/she wants to probe trans-Planckian physics, he/she 
  cannot avoid localizing the energy $L$ within the interval $1/L$, with all the above-considered consequences. 
  
   \section{Influence of a  Possible Black Hole Information Loss} 
 
      The question we would like to address is,  whether our reasoning about deep-UV-completeness of Einstein gravity  is sensitive to the possible information loss by a BH \cite{paradox}. 
      The answer to this question is negative, as we shall now explain. 
      
       As discussed above, the deep-UV-completeness of Einstein gravity follows from the impossibility 
   of probing distances $L \, \ll \, L_P$.    
        The reason for this barrier, as we have explained, is in the  fact that in order to probe a distance 
        $L$, one has to pump energy $E = 1/L$ within that distance.  But, for 
        $L\, \ll \, L_P$  the Schwarzschild radius of this localized portion of energy is much larger 
     than the  Planck length. As a result, any such attempt  will end up by a formation of 
     a BH with horizon $R\, = \, L_P^2/L$,  way before one can approach the sub-Planckian distances.  Thus, the entire information extracted from such a measurement will be restricted 
     by the information encoded at the horizon of a resulting classical BH.  
   
  We now wish to make the following two comments. 
  
  First,  regardless whether the subsequent evaporation of 
 a classical BH violates information or not, its formation does represent the insuperable barrier for the  short distance measurements. 
 
 Secondly,  since the dynamics of a large semi-classical BH 
 is  governed by IR Einsteinian gravity, any inconsistency (e.g., such as information loss, or violation of unitarity) would signal the incompleteness of Einstein gravity in IR, rather than in UV.  
   
   Our assumptions exclude such an inconsistency.   We rely on the fact that pure Einstein 
   (super)gravity  is a consistent theory in IR.  Existence of any inconsistency in IR  would mean, that new light degrees of freedom must be integrated in, which would contradict to our starting 
   point that the only propagating IR degree of freedom is the massless graviton.  
   
     Finally,  since we are working in pure Einstein gravity without any non-gravitational species,  the only information encoded in the BH can be  in the form of gravitons.  Discussions about the BH information loss, typically involve other probe states (e.g., such as fermions with baryon number). 
  As we shall see, this seemingly innocent deformation of the theory dramatically affects its properties. In particular,  in such a case existence of extra propagating gravitational degrees of freedom is a must, and analysis has to be changed accordingly.

  \section{Large Distance Effects of Trans-Planckian States}
  
   We wish to discuss, why trans-Planckian physics cannot have any observable  long-distance effects that could show up in very precise measurement. 
   For example, why their influence could not  modify the metric of gravitating sources, and for instance,  affect the dynamics of BH formation.  
     The reason why long-distance measurements cannot establish any contact with  trans-Planckian  physics, is again rooted in the fact that trans-Planckian degrees of freedom cannot be perturbative quantum states.  The only physical meaning they can carry is of the macroscopic classical object that  belong to the deep IR region of the theory.   In this respect,  any trans-Planckian state is not any better probe of UV physics than any other macroscopic BH of the same mass. 
     
    In order to make this discussion more concrete, we shall first consider a simplified toy model with two  scalar ``gravitons", which crudely captures the essence of the phenomenon. 
   Let us consider a theory with the graviton that propagates two spin-0 degrees of freedom. One, call it $\chi$, will be assumed to be massless and will be the analog of Einstein graviton. The other one, $\phi$, will be a heavy state with the mass $m$, which at the beginning we shall take below 
   $M_P$ and later push into the trans-Planckian region, $m \, \gg \, M_P$. 
 The two degrees of freedom couple to the energy momentum sources through an effective  metric,  
 \begin{equation}
 g_{\mu\nu} \, = \, \eta_{\mu\nu} \,  + \, \eta_{\mu\nu} (\chi \, +\, \phi)/M_P \, .
 \label{fmetric}
 \end{equation}
   We wish to study the long distance corrections to the metric produced by a heavy source $T_{\mu\nu}$.   For our purposes it will be enough to  work up to the second order in $G_N$.   Therefore, we restrict ourselves by considering up to 
   trilinear couplings of the gravitons. 
  As said above, we shall first keep the mass of the heavy graviton below $M_P$, and later take trans-Planckian limit.    
    The Lagrangian is: 
  \begin{equation}
  (\partial_{\mu} \chi)^2 \, +   (\partial_{\mu} \phi)^2 \, - \, m^2  \phi^{2}  \, + 
  {1 \over M_P}  (\chi\, + \, \phi) (\partial_{\mu} \chi)^2 \, + \  {1 \over M_P}  (\chi\, + \, \phi) \, T\, , 
  \label{twoscalars}
  \end{equation}
  with the corresponding equations of motion,
  \begin{equation}
  \partial^{\mu} ((1\, + \, 
  {1 \over M_P}  (\chi\, + \, \phi)) \partial_{\mu} \chi) \,
  - \, {1 \over 2 M_P} \, (\partial_{\mu} \chi)^2\, = \,  {T \over 2M_P}\, ,
  \label{phione}
  \end{equation}
  and 
  \begin{equation}
 \square\, \phi\, + \, m^2 \, \phi \,  - \, {1 \over 2 M_P} \, (\partial_{\mu} \chi)^2\, = \,  {T \over 2M_P}\, .
  \label{phione}
  \end{equation}
   Let us evaluate the above system for a static localized source of mass $M$.  Since we are interested in the metric outside the source, the latter can be approximated by $T \, = \, 8\pi \delta(r) M$. 
  In the linear order the two gravitons contribute into the metric  as, 
  \begin{equation}
  {\chi^{(1)} \over M_P}  \, = \, {R \over r} ~~~ {\rm and }~~~{\phi^{(1)} \over M_P}  \, = \, {R \over r} \, ,{\rm e}^{-mr} \, .
  \label{linearscalars}
  \end{equation}
  where  $R \, \equiv \, M/M_P^2$ is the gravitational radius of the source. 
   Thus, to the linear order,  the correction to the metric from the heavy scalar relative to the massless one  is exponentially small.   The stronger relative correction occures in the next order 
   in $R/r$.  In the diagrammatic language this corresponds to a Feynman diagram with a cubic vertex from which  the two graviton lines are ending on the source. 
   As it is easily checked from the equations, 
 the second order correction to the  metric are, 
 \begin{equation}
 {\chi^{(2)}  \over M_P} \, \propto \,  {R^2 \over r^2} \, 
 ~~~~{\rm and}~~~ {\phi^{(2)}  \over M_P} \, \propto \,   {R^2 \over r^2} {1 \over (rm)^2} \,.
 \label{secondorder}
 \end{equation}
  
    We see that unlike linear order, in the second order the relative correction from the heavy state is suppressed only by the power $(mr)^{-2}$.  This fact can be understood as  the correction to the non-linear 
coupling of the massless graviton $\chi$ to the source, due to the exchange of the heavy state $\phi$. 
The reason, why this effective interaction is not exponentially suppressed is because the  virtual 
heavy state does not have to propagate distances larger than its inverse mass. 
Indeed, if we explicitly integrate out the heavy scalar, we will induce the following effective coupling between the 
massless graviton and the source 
\begin{equation}  
  {(\partial_{\mu} \chi)^2 \over M_P^2m^2} \,T  \, .
  \label{newcoupling}
  \end{equation}
  
    The following two points emerge from the above consideration. 
    
  First, as  long as $m \,  < \, M_P$,  the corrections to the metric coming from the heavy state 
  is suppressed  at least by the powers of $(mr)^{-1}$. This implies that the heavy state
 cannot interfere with  gravitational processes at distances larger than $m^{-1}$. For example, formation of a BH of the gravitational radius $R\, \gg \, m^{-1}$ will not be affected. 
  
   Notice that the gravitational radius in Einstein theory can precisely be deduced from 
  equating  the leading and subleading contributions.  That is,  we approach the ``horizon" when $\chi^{(1)} \, \sim  \, \chi^{(2)}$.   At the horizon all higher order corrections become equally important, and the series have to be re-summed.  
       
   Secondly, for $m \,  < \, M_P$, one can argue,  that although the fact of large BH formation is unaffected, the deviation from the Einsteinian dynamics should be observable by precise measurements of the corrections to the metric.  In other words, by measuring  corrections of order $\, \sim \, 1/(mr)^2$ to the metric , we can deduce the information about the heavy physics.  This is certainly true as long as the mass $m \, \lesssim \, M_P$, but
 for trans-Planckian states the situation changes dramatically.   
  
    Again, the reason is,  that as soon as we push the mass of the $\phi$ into the trans-Planckian 
    region, $m \, \gg \, M_P$, $\phi$ stops to be a quantum particle and becomes a classical object, and this must be taken into the account. For instance, the operator  (\ref{newcoupling}) in reality 
 will be exponentially suppressed at least by the entropy factor $e^{-s} \, = \, e^{-m^2/M_P^2}$, but more importantly, it should be indistinguishable from the operator obtained by integrating out any other classical BH of the same mass $m$.  Thus, operators generated by integrating out 
$\phi$, stop revealing information about the length scale $m^{-1}$ as soon as the latter becomes shorter than $L_P$.   Starting from this point, $\phi$ becomes less and less efficient probe of the short-distance physics and only carries information about the scale 
$R_{\phi} \, =\, (L_P^2m)$ rather than $m^{-1}$.

     We can now repeat the same analysis replacing the massless scalar graviton by the real spin-2 Einstein graviton , $h_{\mu\nu}$.    The equation  (\ref{phione}) is now replaced by the 
Einstein equation,
 \begin{equation}
    G_{\mu\nu} \, = \, 8 \pi G_N\, T_{\mu\nu} \, 
    \label{einstein} 
 \end{equation}
  which to the linear order in graviton can be written as 
(in harmonic gauge $\partial^{\mu} h_{\mu\nu} \, = \, {1 \over 2} \partial_{\nu} h$):  
  \begin{equation}
  \square  \, h_{\mu\nu} \, = \, - \, 16\pi \, G_N \, (T_{\mu\nu} \, - \, {1 \over 2} 
  \eta_{\mu\nu} \, T_{\alpha}^{\alpha} \, )
   \label{gravitoneq}
  \end{equation}
  where, $h\, \equiv \, h_{\mu}^{\mu}$. To the linear order in $G_N$ this gives a familiar result,  
 \begin{equation}
 {h^{(1)}_{\mu\nu} \over M_P}  \, = \,  {1 \over M_P^2} \,  {T_{\mu\nu}\, - \, {1 \over 2} 
   \eta_{\mu\nu} T
     \over \square } \, ,
 \label{proplinear}
 \end{equation}
 which for a static point-like source $T_{\mu\nu} \, = \, \delta_{\mu}^0\delta_{\nu}^0 \, M\, \delta(r)$ becomes, 
  \begin{equation}
   {h^{(1)}_{\mu\nu} \over M_P}  \, = \, \delta_{\mu\nu} \, {R \over r} \, .
  \label{lineargraviton}
    \end{equation}
  The horizon corresponds to a distance $r$ for which the above contribution becomes order one.  In the same time, all the higher order (in $G_N$) contributions, by consistency, become equally important, and the series have to be re-summed.   
      
   Diagrammatically, these corrections correspond to the processes when multiple gravitons emitted by the source interact non-linearly.  This is equivalent to solving the Einstein equation in the given order in $G_N$, which effectively takes into the account self-sourcing of graviton 
 by its energy-momentum tensor.  For example, to second order in 
  $h_{\mu\nu}$ we have,  $8\pi G_N \,  T_{\mu\nu} (h) \, = \,  
 - {1\over 2} \, h^{\alpha\beta} \,  
 \partial_{\mu}\partial_{\nu} \,  h_{\alpha\beta} \, + \,  ... $. 

   Evaluating this for the linearized graviton contribution $h^{(1)}_{\mu\nu}$, we  
 get the corrections to the metric in the second order in $G_N$. For example, 
  \begin{equation}
   {h_{00}^{(2)} \over M_P}  \, = \,  {1 \over 2} \, {R^2 \over r^2} \,~~~~
    {h^{(2)} \over M_P}  \, = \, - \,  {1 \over 2} \, {R^2 \over r^2}  \, .   
  \label{bilineargraviton}
    \end{equation}
   Again, these corrections confirm, that the horizon is at $r = R$. Beyond this point, the expansion in $G_N$ is no longer valid and the series have to be re-summed. 
   
   The  effect of the massive scalar graviton $\phi$
 is,  that  $h^{(2)}_{\mu\nu}$ gets corrections also from the coupling to 
 the energy momentum of $\phi$, 
 \begin{equation}
 T_{\mu\nu}(\phi) \, = \, \partial_{\mu}\phi \partial_{\nu}\phi \, - \, {1 \over 2}\, 
 \eta_{\mu\nu}
 (\partial_{\alpha}\phi \partial^{\alpha}\phi\, + \, m^2 \phi^2) \, + ...\, .
 \label{scalarenergy}
 \end{equation}
 This  has to be evaluated on the first oder solution 
 $\phi^{(1)}\, = \, e^{-mr} (R/r)$, and obviously  gives an exponentially-suppressed contributions 
 to $h^{(2)}_{\mu\nu}$. 
 
   A more important,  power-law suppressed, corrections can also appear if there are couplings between $\phi$ and $h$ of the form 
 \begin{equation}
   { \phi  \partial^n h^k  \over M_P^{n+k-3}} \, ,  
   \label{operators}
   \end{equation}
   (gauge invariant contraction of indexes is assumed).  
   Just as in the scalar example case, after integrating out $\phi$, we will induce  corrections to the  effective metric,    
  \begin{equation}
 g_{\mu\nu} \, = \, \eta_{\mu\nu} \, + \, {h_{\mu\nu} \over M_P} \,  + \eta_{\mu\nu}  { (\partial^n h^k)  \over M_P^{n+k-3}} \, + \,  ... \, , 
   \label{metriccorrected}
   \end{equation}
which after being evaluating on the solution for $h$ will give power low corrections 
to the long-distance metric.  Although not playing a significant role in long distance gravity,  
these corrections can certainly be measured and serve as a probe of 
short distance physics, as long as $m < M_P$.  

     However, for $m \, \gg \, M_P$ 
  the new degree of freedom is no longer a perturbative state, but a macroscopic BH, and 
   belongs entirely to large-distance sector of the theory.    
    It becomes a classical  BH of horizon $R_{\phi}\,  = \, m/M_P^2 \,   \gg \, L_P$.  
  
    Again, $\phi$  now has to be considered as a 
    sequence of sources that emit arbitrary number of gravitons that merge in non-linear vertexes.
   These corrections contribute powers of 
  $(R_{\phi}/ r)$ to the metric, which have to be re-summed  at $r \, \sim \, R_{\phi}$.  
  
  This is the diagrammatic indication of the non-perturbative fact that $\phi$ is a BH and develops 
  horizon.  Thus, at this point $\phi$ can no longer be regarded as the propagating quantum 
  degree of freedom, and integration  over $\phi$ has to be performed as the integration over 
  a classical object.   Again, as in the examples considered earlier, the interaction vertex between 
 $\phi$ and other quantum propagating degrees of freedom now has to be understood as an effective vertex  controlling the quantum decay (i.e., Hawking evaporation) of a classical BH into the quantum  particles in question. 
  For example,  any vertex of the form  (\ref{operators}) now describes the evaporation 
of $\phi$-BH into   $k$-number of massless gravitons,  
\begin{equation}
\phi \,   \rightarrow \,   k-{\rm number~ of~gravitons} \, .
\label{inhdecay}
\end{equation}    
 Since this process describes a quantum decay of a semi-classical thermal object of temperature 
 $T \, = \, M_p^2/m$ in  $k$-number of  quantum particles  of energy $m \, \gg \, T$, the  
 rate of this decay must be exponentially-suppressed by the Boltzmann factor 
 e$^{-(m/T)}$.   This suppression factor has to be included in the effective strength of the vertex.    
  
  Thus, in any process in which $\phi$ appears as an internal virtual state, 
  the contribution is exponentially suppressed  at least by e$^{-(R_{\phi}m)}$.    
  At this point, running $\phi$ in a virtual line is no any different than running any other 
  classical BH in the same line. 
   
    To summarize, operators that we can obtain by integrating out $\phi$ cannot be any different 
  from what we would obtain by integrating out an ordinary classical BH of the same mass. 
  In other words, by becoming trans-Planckian, $\phi$ stopped to be a quantum degree of freedom  and  became classical, with the minimal size given by $R_{\phi}$.

 \section{Difference of Gravity from Other Non-Renormalizable Interactions}
 
     From the above reasoning it is clear that because of BH barrier the trans-Planckian region of Einstein gravity is equivalent to the deep-IR region.   The maximal information that can be extracted from any sub-Planckian distance 
  $L$ cannot exceed the information carried by a classical BH of size $L_P^2/L$. 
    In this way, Einstein gravity is  self-UV-complete.     
 
    In order to stress the profound uniqueness of gravity, let us compare it to any other 
non-renormalizable interaction. Consider, for instance, the interaction of the Nambu-Goldstone bosons in  $O(n)$ sigma model, with the following action, 
\begin{equation}
V^2\partial_{\mu} {\mathcal O(x)}^T \partial^{\mu} {\mathcal O(x)}, 
\label{sigmamodel}
\end{equation}
where  ${\mathcal O(x)_a} \, \equiv\,  O(x)_{ab} n_b$ denotes an arbitrary local 
$O(n)$-transformation acting on a constant fundamental  $n$-vector $n_a \, \equiv 
\, (0,0, .....1)$  ( with $a,b \, = \, 1,2,...n$) and $V$ is the scale. 
 The angular degrees of freedom represent Nambu-Goldstone bosons, which after canonical normalization acquire derivative interactions suppressed by the scale $V$.  These derivative interactions are seemingly similar to gravity.   First,  they both decouple at low energies, and  
 become strong at the scale  $V$ above which the perturbative 
 unitarity is violated.  So naively, the scale $V$ for Nambu-Goldstone bosons plays the role which is similar to the one that 
 $M_P$  plays for gravity.   But, the difference between the two cases is fundamental. 
 In contrast with  gravity, in the above theory nothing prevents us from probing distances shorter than the scale $1/V$.   
 
  In order to restore the consistency above the scale $V$, we need to integrate in a new radial degree of freedom, by allowing the 
  absolute length of the unit vector to fluctuate. 
  In the other words we promote the scale $V$ into a vacuum expectation value (VEV) of a fundamental scalar $\Phi(x)$ in the following way, 
  \begin{equation}
 \Phi_a(x)\, = \,  \left (1 \, + \, {\rho(x) \over V} \right ) {\mathcal O}(x)_a\, ,  
  \label{radialmode}
 \end{equation}   
 where $\rho(x)$ is the radial mode. 
        In this way, the $O(N)$ sigma model is promoted into a Nambu-Goldstone model with the
  spontaneously-broken $O(N)$-symmetry at the scale $V$,  
 \begin{equation}
 \partial_{\mu} \Phi(x)_a \, \partial^{\mu}\Phi_a \, - \, \lambda (\Phi_a\Phi_a \, - \, V^2)^2 \, .
 \label{gold}
  \end{equation}
  The field $\Phi_a$ differs from the sigma model field only through the existence of the radial mode
$\rho(x)$.
 
   In sharp difference with gravity, the  existence of the radial mode is crucial for restoring unitarity at all the energies above $V$.  This is because in the case of the sigma-model there 
   is no BH barrier, and physics can be probed down to arbitrarily short lengths.
   
   In case of pure Einstein gravity, even if we introduce some new degrees of freedom right at 
   $M_P$, these will not play any role in restoring consistency of the theory in deep-UV. This role is taken up by the massless graviton. 
   Moreover, we want to stress, that even if we  do not introduce any new degrees of freedom, 
  some quantum particles will nevertheless appear around $M_P$. The existence of such states follows from the fact, that at the very last stage of evaporation 
  BHs are essentially indistinguishable from quantum particles.  But, again, these states play no role in deep-UV.

  \section{Quantum Particles of Planck Scale Mass in Spectrum of Einstein Gravity}
  
    We wish to point out that Hilbert space of Einstein gravity contains quantum particle states with mass $\sim \, M_P$.  Existence of such states is not an additional assumption, but is 
    built-in in Einstein gravity.  Their presence follows from the existence of classical BHs. 
    
      Let us consider Einstein gravity at large distances.  This sector of theory contains classical
      BHs of mass $M \, \gg \, M_P$. 
      The half evaporation time of these objects is given by \,  
   \begin{equation}
   \tau_{BH} \, = \, c \, L_P (ML_P)^3 \, ,
   \label{taubh}
   \end{equation}
   where  $c$ is a numerical constant.   For us,  the important fact is that $c$ is sufficiently larger than one, which implies
   that semi-classical black holes live much longer than their inverse mass. 
    The black holes with the Schwarzschild radius $R \, \gg \, L_P$ are classical objects,  since their Schwarzschild radius exceeds the Compton wave-length $M^{-1}$.  We shall only be interested in BHs that do not carry excessive charges (such as an electric charge) that 
could stabilize them in the classical region. 
   
     Let us parametrically decrease the mass $M$.   Once $R\, \sim \, L_P$, the BH crosses 
     into the quantum region, since its Compton wavelength exceeds the Schwarzschild radius. 
      At this point, the semi-classical description of the BH breaks down, and it has to be treated as a quantum state. Of course, in this regime the eq(\ref{taubh}) is no longer applicable, but 
      by continuity,  the balance between the mass and the decay width should be maintained. 
      In other words, because in the semi-classical domain there is a strong hierarchy between 
      the decay width and the mass, 
      \begin{equation}
      \Gamma_{BH} \, = \, \tau_{BH}^{-1} \, \ll \, M\, ,
      \label{width}
      \end{equation}
 it is parametrically impossible to cross over from the semi-classical long-lived  
   state directly into a quantum broad resonance, without passing an intermediate stage 
   of a sharp quantum resonance. This intermediate state corresponds to a quantum particle 
   of mass $\sim\, M_P$.  To make a more precise estimate  is hard with the current knowledge of 
   properties of the micro BHs. However, for us the important thing is the very fact of existence 
   of such quantum states. 
   
      One may wonder, how robust is the existence of  quantum states around $M_P$. 
      For example, what if evaporating  BHs either never reach the Planck mass,  or cross over and continue existence with masses $\ll M_P$? 
     In fact,  none of the above is possible  in  Einstein gravity in which only propagating degree of freedom below $M_P$ is a massless graviton.          
   
    First,  in  Einstein gravity there always will be BHs that will reach the  $M_P$ mass in their evaporation process. 
    The BHs with size $R\, \gg \, L_P$ are in the classical regime and their properties are 
  well understood.  Such a BH can only stop evaporation if it becomes an extremal state. 
  That is, its charge $Q$ (under some gauge symmetry that must be the part of the IR sector of the theory) has to become equal to its mass measured in  $M_P$-units.   Thus,  only the BHs with 
  excessive charge ($Q \,  \gg \, 1 $) can be stabilized in trans-Planckian mass region, 
  $M\, = \, Q\, M_P\, \gg \, M_P$.   Notice, that this charge must be pre-existing, and cannot be acquired in the evaporation process. 
   This is because the neutral semi-classical BHs evaporate democratically in particles and anti-particles and thus cannot accumulate any net charge in the evaporation process.  
     Thus, any BH of sufficiently small charge is bound to shrink down to the Planck size. 
     
     Now let us ask if a BH could cross over and continue existence  with mass $M \, \ll \, M_P$. 
     This can only happen, if in the IR region of the theory there is a quantum state to which the BH can evolve.  For example, an electron can easily 
  be an end result of evaporation  a BH of unit electric charge and  $1/2$ spin, but this is a triviality, since the corresponding quantum state, the electron,  is already part of the IR spectrum of the theory. 
  In other words,  the end results of the BH evaporation cannot add any new state to the IR region of the theory.  Thus,  in pure Einstein gravity the new quantum states appear only around $M_P$.

    The reason why we were able to deduce the existence of  the above quantum states in pure Einstein gravity is the UV-IR connection. This connection follows from the fact that any perturbative degree of freedom whenever  pushed into the trans-Planckian region  bounces back 
    to an IR sector of the theory in form of a classical  BH. 
    
      In other words, in gravity,  although heavy states decouple from the quantum processes, 
      they inevitably become part of the classical IR sector of the theory. 
      
      In contrast, in ordinary field theories the heavy perturbative states simply decouple without 
    having any IR counterparts. 
   
     In order to understand this profound difference, consider the example of an 
     $O(3)$ sigma model described by the action (\ref{sigmamodel}).   In this theory there are 
     no classical object that indicate existence of new degrees of freedom at the scale $V$.
     Of course, there are classical solutions, but they do not correspond to a  
    massive limit of  perturbative states.   Here we are not talking about correspondence between the solitons and perturbative states in terms of electric/magnetic type duality. We are interested in perturbative and classical states that cross to each other within the same weakly-coupled description. 
     
        For instance, the $O(3)$ sigma model admits a monopole solution, 
        \begin{equation} 
         n_a \, = \, {x_a \over r}  \, ,
       \label{monopole}
       \end{equation}
     where $x_a$ are cartesian coordinates, and $r$ is the  radial one.    
   But, monopole carries no information about existence of any massive quantum degree of freedom  around the scale $V$.    
           
        The same is true in the gauged version of the theory.   Gauging of non-linearly realized $O(3)$ symmetry introduces a massive $W_{\mu}$-boson and a massless  photon. 
    The configuration (\ref{monopole}) now acquires an $U(1)$-magnetic charge, and a mass 
 $M_{mon} \, = \, {M_W  \over g^2}$, but again, this classical solution
 carries no information about the existence of extra degrees of freedom. For instance, from the existence of the magnetic monopole of the finite mass, we can certainly deduce the existence of the $W$-bosons, but this is only because both the monopole and the $W$ boson probe exactly the same length-scale, the Compton wave-length of the
$W$-boson.  The monopole is just a coherent state of $W$-bosons. 
      However,  from the existence of monopole,  we cannot deduce the existence of  a  heavy Higgs particle (\ref{radialmode}).  
          
     The reason is that Higgs particle decouples without leaving any classical trace in IR.          
  If we make it arbitrarily heavy,  Higgs will probe arbitrarily short distances, and will decouple 
  from classical IR.       
    
         Such a decoupling is impossible in gravity, since by making a quantum state heavy, we will eventually make it classical and vice versa, by making a classical BH lighter we will sooner or later cross in the quantum regime. 
      Because of this fundamental property, at the crossover of the two domains the presence 
      of heavy quantum species is inevitable. These  are the states with masses $\sim M_P$,  existence of which is built-in in Einstein gravity.

 \section{Beyond Einstein: Generalizing the Notion of the Planck Length}. 
 
   We now wish to extend our analysis to the theories that include new gravitational degrees of freedom on top of the Einsteinian massless graviton, and derive a criterion of deep-UV completion of such theories by the BH barrier. 
   
    \subsection{What is Gravity?}
    
   Quantum field theories are fully characterized by the propagating degrees of freedom and by their  interactions.   Here we shall be interested in theories that include only gravitational degrees 
   of freedom.  We shall define the latter as the propagating degrees of freedom that  are sourced 
   by the conserved energy momentum tensor  $T_{\mu\nu}$ and which in the classical limit 
   and for localized sources amount to the local deformations of an asymptotically-flat metric 
   $g_{\mu\nu}\, = \, \eta_{\mu\nu}  \, + \, \bar{h}_{\mu\nu}$. 
   By conservation of the source, to the linear order these include only spin-0 and spin-2 fields. 
    We shall therefore restrict ourselves by considering theories with maximal spin equal to $2$.  
  Thus, a small metric perturbation around the flat background ($\bar{h}_{\mu\nu}$) in such theories will contain not only the massless spin-2 state,  $h_{\mu\nu}$,   but also an arbitrary number of massive spin-2 ($h_{\mu\nu}^{i}$) and spin-0 ($\phi_{\mu\nu}^{j}$) fields: 
  \begin{equation}
 \bar{h}_{\mu\nu} \, = \, {1 \over M_P}  \left (h_{\mu\nu}\, + \, \sum_i \, c_2(i)  \, h_{\mu\nu}^{i}\, + \,  \sum_j \, c_0(j) \, \phi_{\mu\nu}^{j}\, \right)  \, .
  \label{spectrum}
  \end{equation}
 Possible spin-1 fields cannot couple to the conserved sources to the linear order, and therefore do not appear 
 in the above decomposition. However, they can participate in non-linear interactions.      
    
     The one-graviton exchange amplitude among the two sources $T_{\mu\nu}$ and $t_{\mu\nu}$ is highly restrictive and takes the following form, 
     \begin{equation}
    T^{\mu\nu} \, \langle \bar{h}_{\mu\nu}\bar{h}_{\alpha\beta} \rangle \,   t^{\alpha\beta}\, =  \, 
   {1 \over M_P^2} \, \left ( {T_{\mu\nu}t^{\mu\nu} \, - \, {1 \over 2} 
   T_{\mu}^{\mu} t_{\nu}^{\nu}
     \over p^2 } \, 
      + \,  \sum_{i} \, \rho_2(i) {T_{\mu\nu}t^{\mu\nu} \, - \, {1 \over 3} 
   T_{\mu}^{\mu} t_{\nu}^{\nu}
     \over p^2 \, + \, m^2_i} \,
     + \, \sum_{j}  \, \rho_0(j) \, {
   T_{\mu}^{\mu} t_{\nu}^{\nu}
     \over p^2 \, + \, m^2_j} \right ) \, ,  
     \label{spectral}
   \end{equation}
   where we have separated the contributions from massless spin-2, massive spin-2 and 
   spin-0 poles.  We have normalized the relative strengths to the one of a zero mode graviton.
   In case of continuum, the discrete sum has to be replaced by the integral.  
   The crucial point is, that all the spectral densities  
   $\rho_2(i) \, \equiv \, |c_2(i)|^2 $ and $\rho_0(j) \, \equiv \, |c_0(j)|^2 $ must be  semi-positive, 
  \begin{equation}
  \rho_2(i) \, >  \, 0, ~~~ \rho_0(j) \, > \, 0 \, ,
  \label{positive}
  \end{equation} 
  in order for the theory to be ghost-free. 
  This structure uniquely defines the linearized metric produced by an arbitrary source. 
  For instance, the gravitational potential produced by a localized non-relativistic mass 
  $T_{\mu\nu} \, =\, \delta_{\mu}^0\delta_{\nu}^0 \, M\, \delta (r)$ is give by,
    \begin{equation}
    \bar{h}_{00}(r) \, =  \, 
   {M \over M_P^2 \, r} \, \left ( {1 \over 2} \, 
      + \,  {2\over 3} \, \sum_{i} \, \rho_2(m_i) {\rm e}^{-m_i r}  \, +
\,    \sum_{j}  \, \rho_0(m_j) \,  {\rm e}^{-m_j r} \right ) \, .  
  \label{potential}
    \end{equation}
    
      What would be an analog of $L_P$ as of the shortest observable distance in such a theory?  
  Without the knowledge of non-linear interactions it is impossible to answer this question.  
  However, we shall formulate a sufficient condition for the existence of such a length scale  in 
  terms of the strong coupling scale  of gravitational degrees of freedom.   
  Let $\Lambda_{str}$ be a lowest energy scale at which some of the gravitational degrees of freedom become strongly-coupled.  Notice, that due to the very constrained tensorial structure 
(\ref{spectral})  and the positive-definiteness of the  spectral functions (\ref{positive}),
the strength of linearized gravity can only grow at short scales \cite{stronggia}.  Due to this, 
$\Lambda_{str}  \, =  \, M_P$ is the upper bound \cite{bdv}, since even if all the massive degrees of freedom remain weakly-coupled, the Einsteinian massless graviton becomes strongly coupled at the Planck energies.  So let us consider the case 
$\Lambda_{str} \, < \, M_P$.  This means that  (some) interactions become strong  at 
the scale $\Lambda_{srt}$, so that in naive perturbative approach, theory requires an UV completion at the distances $L \, \ll \, 1/\Lambda_{str}$.   So let us ask the question, under what circumstances the BH barrier shields such distances and UV-completes the theory? 

 The criterion can be formulated in terms of the BH properties in the following way. 
 First, since we are interested in theories that in deep-IR flow to Einsteinian gravity with  the only 
 propagating degree of freedom being a massless spin-2, we shall assume the existence of a mass gap $M_c \, \equiv \, 1/R_c$  corresponding to a first massive excitation in expansions
 (\ref{spectrum}) and (\ref{spectral}). Later, in some examples, we shall consistently take 
 the continuum limit, $R_c \, \rightarrow \, \infty$, but for a moment we shall keep 
 the gap finite. 
 
  We thus have a hierarchy of scales, 
  \begin{equation}
   L_P \,  \lesssim \, 1/\Lambda_{str}\, \, \lesssim \, R_c \, .
 \label{scales1}
   \end{equation} 
 In such a theory, at distances $r\, \gg \, R_c$, the only propagating degree of freedom is 
 a massless graviton and gravity is pure Einsteinian, with all the usual properties. In particular, 
 in such a theory there must exist Schwarzschild BHs of radius $R \, \gg \, R_c$. 
 The mass-to-radius dependence for such a BH is given by the usual Einsteinian relation,
 $R(M) \, = \, ML_P^2$. Now, since $R\, \gg \, L_P$, we have $R \, \gg \, M^{-1}$. 
 In other words, by taking $M$ large, the Schwarzschild radius of such a BH can be made  arbitrarily larger than its Compton wavelength.  
 
  Now, let us start decreasing $M$ parametrically.  Of course, for $R(M) \, > \, R_c$ we are in Einstein's gravity and $R(M)$ decreases linearly with $M$.  Once $R(M)$ drops below 
  $R_c$, we are no longer in Einsteinian regime and dependence of $R(M)$ on 
  $M$ can change, in general. 
  
   If $R(M)$ and $M^{-1}$ meet for some $M=M_*$, then the (first) meeting point,
   \begin{equation}   
  R(M_*) \, = \, M_*^{-1}, 
   \label{meetingpoint}
   \end{equation}   
 marks the start of the BH barrier. The corresponding length scale $L_*\, \equiv \, M_*^{-1}$
   is the shortest observable length of nature.  This scale plays the same role as the Planck length plays in Einstein gravity. 
   
     The measurement attempted at any shorter 
   scale $L \ll L_*$, will result into the formation of a classical BH of size $R(1/L) \, > \, L_*$. 
   For example,  in Einstein gravity, the relation is $R(M) \, = \, ML_P^2$ and the meeting point is $M\, = \, M_P$.

 
  Thus,  the whole issue,  when is gravity able to UV-complete a strongly-coupled  interaction, 
 is reduced  to the question, whether $R(M)$ and $M^{-1}$ meet before 
 $R(M)$ crosses with $1/\Lambda_{str}$.   That is, whether  $M_* \, < \, \Lambda_{str}$. 
     Obviously, the physically-observable strong coupling scale $\Lambda_{str}$
 can only be at or below the meeting point $M_*$, but never above. 
  The meeting point marks the beginning of the BH barrier, and any coupling 
  that formally gets strong above this energy is shielded by the BH physics.  
 
     On the other hand, if  $M_*\, \gg \, \Lambda_{str}$,  
  interactions become strongly-coupled way before the BH barrier can interfere and restore consistency. In such a case, 
  theory requires an independent UV-completion above the scale $\Lambda_{str}$. 
  
    To summarize, the BH barrier restores consistency in a strongly-coupled theory 
  as long as, 
  \begin{equation}
    M_* \,  <  \, \Lambda_{str}  
   \label{star} 
    \end{equation}
      

    
     This criterion can be re-formulated in terms of holography.  For this, in any theory
   that satisfies standard energy-positivity conditions   
       we can define a shortest length scale  $L_*$ on which one can store (and retrieve)  information. 
  Since the storage of a single information bit on a scale $L$ costs energy $1/L$, 
  the scale $L_*$ is the minimal wavelength that exceeds its  Schwarzschild  
  radius, which is the same as (\ref{meetingpoint}). 
  In Einstein  $L_*\, = \, L_P$.  Now whenever in a given theory 
     $\Lambda_{str}^{-1}  \, \gg \, L_*$, such theory requires UV completion at the scale $\Lambda_{str}$ and cannot be  saved by BHs.

     Let us now consider some examples. 
     
     \subsection{Kaluza-Klein Theory} 
     
    An example of UV complete gravity that satisfies relation (\ref{star})  is provided by  Kaluza-Klein theories. 
Consider for example a $5$-dimensional theory compactified 
    on a circle of radius $R_c$.  As it is well-known, from the $4$-dimensional point of view this is a theory of the tower of massive spin-2 states.  The  spectral decomposition (\ref{spectral}) in this case takes the form, 
    \begin{equation}
    T^{\mu\nu} \, \langle \bar{h}_{\mu\nu}\bar{h}_{\alpha\beta} \rangle \,   t^{\alpha\beta}\, =  \, 
   {1 \over M_P^2} \,   \sum_{n} \,  {T_{\mu\nu}t^{\mu\nu} \, - \, {1 \over 3} 
   T_{\mu}^{\mu} t_{\nu}^{\nu}
     \over p^2 \, + \, n^2/R_c^2}  \, .  
     \label{kk}
   \end{equation}
    Although each KK graviton couples by $1/M_P$ suppressed interaction, because of multiplicity the strong coupling {\it universally} happens at the scale $\Lambda_{str}^3\, = \, M_P^2/R_c$, which is simply a $5$-dimensional Planck mass. The same scale sets the crossing point for the 
    BH Schwarzschild radius  $R(M)^2 \, = \, M\, R_c/M_P^2$ and its inverse mass.
   Theory contains no other strong coupling scale, and because of this,  distances 
   shorter than $L_5 \, \equiv \, (R_c/M_P^2)^{-1/3}$ cannot be probed, in principle. 
 Theory is deep-UV-complete.   This result is not surprising since in deep-UV the KK theory is just a five dimensional pure-Einstein gravity.  
 
   Similar properties must be shared by pure supergravity theories in $D$-dimensions since by supersymmetry  the only strong coupling scale for all degrees of freedom is a $D$-dimensional 
   Planck scale.

 \subsection{Examples Not  UV-Completed by Black Hole Barrier}
 
   We shall now consider examples that cannot be UV-completed by Einsteinian BHs,  and 
 require some additional physics in order to restore consistency at short distances. 
   Essentially any physics for which  the probe of strong coupling is not accompanied by BH formation serves as such an example. 
   
     For instance, consider a non-linear  sigma model with the scale $\Lambda_{str} \, \ll \, M_P$ coupled to Einsteinian gravity.   The ordinary pions coupled to gravity would serve as simplest realistic example of this sort.  Obviously, pion interactions are getting strong at energies 
  above the pion decay constant $f_{\pi} \, \sim  $GeV, and theory requires UV-completion at distances $\ll \, f_{\pi}^{-1}$. However, Einsteinian gravity cannot provide such an 
  UV-completion, since  distances $\ll \, L_{\pi}$ can be probed without encountering any BH barrier. 
 Thus, such a theory requires an independent physics for UV-consistency, and 
 as we know, QCD provides one.   
  
    Another example of the same sort is given by the electroweak non-abelian gauge field with a hard mass,  $M_W$. 
  As it is well known, scattering of longitudinal $W$-bosons becomes strong at energies
  above  $M_W/g_W$, where  $g_W$ is the gauge coupling.    Again, for   $M_W/g_W \, 
  \ll \, M_P$, such a theory cannot be UV-completed by Einsteinian gravity, and requires 
  additional physics, such as the Higgs field (however, see the discussion in the outlook section,  about possibility of completing by KK gravity).  
  
   In both above examples, the UV-incomplete physics comes from non-gravitational dynamics. 
   We shall come back to this issue in more details later. Now we wish to provide an example 
   in which UV-incomplete physics comes from  gravitational degrees of freedom. That is, the degrees of freedom sourced by the energy momentum tensor. 
   
   An example of the above sort is a theory which on a Minkowski background  together with 
     Einstein's graviton $h_{\mu\nu}$ propagates an additional scalar, $\phi$.  The non-linear Interactions of graviton are fixed by the general covariance and are controlled by $M_P$. 
      However, let us assume that the scalar possesses a self-coupling of the following sort, 
    \begin{equation}
  {1 \over \Lambda_{str}^3}  \square \phi (\partial\phi)^2  \, ,  
   \label{trilinear}
   \end{equation}
  where  $\Lambda_{str}$ is some scale that can be taken arbitrarily smaller than $M_P$.   The scalar of the above sort appears in  the model of \cite{dgp}.  However, the model we are considering now  
is not this  theory, which would be much harder to analyze in the present context. 
 Rather, the above theory is a simplified prototype, which is substantially different.  The difference  is, that in the original model of \cite{dgp} $\phi$ is not an independent scalar, but rather a helicity-zero polarization of a massive spin-2.  The different helicities decouple only in a very special limit\cite{prl, nicol}, in which $M_P$ is taken to infinity.  Since we don't want to take such a limit, and moreover we wish to keep 
 the graviton massless, we therefore introduce $\phi$ as an independent field, coupled to Einstein gravity. The only similarity we borrow from the model of \cite{dgp}  is  the self-coupling (\ref{trilinear}) \cite{ddgv}.  The reason why we choose such a form of the interaction is,  that on one hand it is becoming strong at the scale 
 $\Lambda_{str}$, and on the other hand theory is ghost-free on the Minkowski background. 
 Thus, such a theory in IR seems to be perfectly consistent.  In order to see what is happening in UV, we  need to recall few properties of the classical solutions. 
 
   Notice,  that in such a theory a static localized source of  mass $M$, produces two types of gravitational radii. First is 
  the usual Schwarzschild radius  $R (M)\, = \,M L_P^2$, which in case of a BH marks the horizon.  
 But in addition\cite{ddgv, andrei}  there is a second scale, the so-called Vainshtein radius\cite{arkady},
  \begin{equation}
   R_V (M)\, = \, \Lambda_{str}^{-1} \, \left ({R(M) \over L_P }\right )^{1/3} \, .
  \label{vainshtein}
  \end{equation}
   The physical meaning of the above radius can be read  from the sphericaly-symmetric solution for  $\phi$, which in the limit  of the decoupled Einstein gravity can be found exactly and takes the form \cite{nicol}, 
 \begin{equation}
 \partial_r \phi(r) \, = \, {\Lambda_{str}^3 \over 4 r} \, \left (\sqrt{9r^4 \, + \, {1 \over 2\pi} R_V^3\, r} \, - \, 
  3 r^2 \right ) \, .
 \label{solution}
 \end{equation}
From this expression it is clear that  Vainshtein radius marks the place where non-linearities in  $\phi$ become important.   For $r \, \gg \, R_V$,  we have $\phi(r) \, \propto \, R/r$, whereas 
for $r \, \ll \, R_V$ the non-linear interaction takes over and we have $\phi(r)\, \propto\, 
\Lambda_{str}^3 R_V^{3/2} \sqrt{r}$.   Crudely speaking,   $R_V$ plays the role for $\phi$ somewhat analogous 
to  the Schwarzschild  radius for Einstein graviton.  

However,  there is a crucial difference.  The Vainshtein's radius {\it is not} a horizon.  So information can be readily retrieved 
from beyond  the $R_V$-sphere,  without encountering any obstacle.   This is the source of the problem, since the strongly coupled region can be experimentally probed and is not protected by the BH barrier. 
 
   Indeed,  we can perform a scattering experiment  that probes distances  $\ll \, 1/\Lambda_{str}$. 
   Again, we have to localize energy $\sim \,  \Lambda_{str}$, within the distance $1/\Lambda_{str}$. 
   The Schwarzschild and Vainshtein radii of this localized energy are 
   $R(\Lambda_{str}) \, = \, \Lambda_{str} \, L_P^2$ and $R_V (\Lambda_{str}) \, = \, \Lambda_{str}^{-1} \, (\Lambda_{str} L_P)^{1/3}$ respectively.  So we see, that both radii are much smaller than the size of the region in which the probe energy is spread, 
   \begin{equation}
  R(\Lambda_{str}) \, \ll  \, \Lambda_{str}^{-1} \, . 
 \label{radii3}
 \end{equation}
 Thus,  Einsteinian BHs are powerless in preventing the access to the distances $\ll \, \Lambda_{str}$. 
 As a result, the theory as it stands is strongly-coupled at such energies and requires UV completion by some non-Einsteinian physics.  
 
  Before abandoning the above example, we wish to stress the following subtlety.
  In concluding that the above theory requires UV-completion,  we were assuming that distances $\ll \Lambda_{str}^{-1}$ can be probed by sources external to $\phi$. What, happens if such 
  probes are forbidden, that is if we decouple $\phi$ from all the other degrees of freedom, is a separate question and requires an independent investigation.

 
   

    \section{Gravity and Species}
    
      We have argued that Einsteinian (super)gravity and its KK extensions are  self-complete in deep-UV.  Why is this not an end of the story?
      In fact, the reason why we cannot declare the success in producing a realistic model of  UV-complete theory of gravity is the existence of {non-gravitational} particle species 
      with the mass much below $M_P$.  Such are the species of the Standard Model, and the problem is their consistent coupling to gravity.  It turns out,  that interactions of 
species with gravity dramatically affects gravitational dynamics \cite{bound, giacesar, bdv}.  We shall now discuss this phenomenon. 


  Consider  Einstein gravity in  $D$ space-time dimensions, in which gravitational interaction is mediated by a $D$-dimensional massless particle of spin-2, the graviton $h_{\mu\nu}$. 
  The corresponding 
 $D$-dimensional Planck mass we shall denote by $M_D$.    As we have discussed, the shortest observable distance  in such a theory is given by $D$-dimensional Planck length $L_D \, \equiv \, M_D^{-1}$, and this fact  makes the theory self-complete in deep-UV. 
  Let us now try to couple this theory to $N$ particle species.  For simplicity we shall take 
  the species to be light.    As we shall see, this seemingly-innocent deformation of the theory 
  dramatically affects the gravitational dynamics.  In fact,  in the presence of light species, it is no longer consistent to assume that gravity is mediated by a single  massless
  graviton, but rather the new gravitational degrees of freedom {\it must be introduced necessarily} \cite{bound, giacesar, bdv}. 
   These degrees of freedom are necessary in order to UV-complete theory at the new fundamental scale, $L_N$,  which is {\it larger}  than  the $D$-dimensional Planck  length  $L_D$,  
 \begin{equation}
 L_N \, \equiv \, N^{1 \over D-2} \, L_D \, .
\label{bound}
\end{equation}
We shall refer to $L_N$ as the {\it species scale}. 

Effect of species on gravity was also considered in some perturbative\cite{other1} and cosmological \cite{other2} contexts. 

      For the detailed proofs  we refer the reader to the original papers \cite{bound,giacesar,bdv}. Here we shall reproduce the argument of \cite{giacesar}.

 This argument is based on impossibility of resolving species identities at the length scales shorter  than $L_N$.  This  is a fundamental obstacle created by gravity. 
    For a  physicists,  existence of $N$ distinct particle species means that he/she can label them, at least in principle, and further distinguish these labels by physical measurements.   Let $\Phi_j$ be the particle species and 
  $ j \, = 1,2, ...N$ be their labels.   Let us now show, that resolving these labels at distances 
  beyond $L_N$ is fundamentally impossible.   The reasoning is similar  to why one cannot 
  resolve distances beyond $L_P$ in theory without species,  with the difference that the existence of species forces this minimal length to grow. 
  
  Indeed,  any process of  decoding the label of an unknown particle  localized within the space-time box of 
  size $L$,  involves comparing the  unknown particle with all $N$ different sample species. 
  Thus,  any such measurement must involve localization of $N$ various species (or of the equivalent information) within the same box.

   This fact automatically limits the size of the box from below. 
 Indeed, localization of each sample particle within the size $L$ costs energy $E\, = \, 1/L$, which implies that the total energy localized  within the region $L$ is at least $E_{Total} \, = \, N/L$.  The corresponding  Schwarzschild radius then is,  
\begin{equation}
  R(E_{Total})^{D-3} \, = \, {N \over L}\, L_{D}^{D-2} \, .
  \label{rpixel}
  \end{equation}
 The key point is,  that species measurement scale  $L$ cannot be decreased arbitrarily,   since eventually its Schwarzschild radius will exceed its own size, and  localized species can no longer be resolved. Any further attempt to decease $L$ will result into the creation of an even bigger BH.  
  The critical size, beyond which the resolution of species is no longer possible is readily derived by  equating the Schwarzschild radius to the the localization size $L$,  
 \begin{equation}
 L^{D-3} \, = \, R^{D-3} \, = \, {N\over L} L_P^{D-2} \, ,  
  \label{lcritical}
  \end{equation}
  which gives the bound on $L$ set by $L_N$.  This constraint is very powerful. It tells us that 
  gravity makes it impossible to resolve the species beyond the scale $L_N$, which means that 
  the new gravitational degrees of freedom must enter into the game. 
  
   So far, we did not  specify the nature of species.  Now it is time to discuss this issue. 
   
   \subsection{KK Species}
   
   The simplest case is when species have just right quantum numbers to fill the KK tower of 
   new  $d$ compact flat dimensions of radius $R_c$.  In such a case everything falls nicely into the places and  relation between $L_N$ and $L_D$  exactly reproduces the usual geometric relation between the $D+d$-dimensional ($L_{D+d}$)  and  $D$-dimensional Planck lengths,
 \begin{equation}
 L_N \, = \, L_{D+d}\, = \, L_{D} \, (R_c/L_{D+d})^d  \, .
 \label{planckDd} 
   \end{equation}
  The reason why $L_N$ came out equal to the $L_{D+d}$, becomes immediately clear if we notice that $N \, = \, (R_c/L_{D+d})^d$ is simply the number of KK species.  It is obvious, that the fundamental length scale of the theory is $L_{D+d}$, and this is precisely what species counting  tells us. 
  As discussed previously,  this theory has a single strong coupling scale, and is self-complete 
  in deep-UV. 
    
     \subsection{Non-Gravitational Species.}
  
      We shall now discuss the effect  of non-gravitational species.    Under this name, we shall 
      refer to the light particles,  with the quantum numbers and interactions that do not fill gravitational (super)multiplets. 
      For instance,  such are the Standard Model particles that cannot be identified with  the fragments of pure high-dimensional supergravity compactified on a smooth manifold. 
      For simplicity, let us consider theory with $N$ massless real scalar fields  $\phi_j$,  coupled to  Einsteinian gravity in $4$-dimensions, with the action, 
    \begin{equation}
     \int \, d^4x \, \sqrt{-g} \,  g^{\mu\nu} \partial_{\mu}\phi_j \partial_{\nu}\phi_j \, .
    \label{scalaraction}
      \end{equation}
      We do not put any specific requirement on the mass  and interaction terms of the scalars, as long as they are light  (with masses $\ll \, L_N^{-1}\, = \, M_P/\sqrt{N}$)  and weakly coupled.  For instance, 
      scalars that inter-couple only through gravity would be sufficient.     
      The above action describes a theory in which below Planck energy the only propagating degrees of freedom are the massless Einsteinian graviton $h_{\mu\nu}$ plus 
      $N$ light weakly-interacting scalars.   
      Seemingly, there is nothing wrong 
      in considering the above action as an effective low energy theory below $M_P$ energies. 
      However, as we know from the previous analysis, this theory as it stands is inconsistent. 
 Namely, it is impossible to avoid the existence of the additional propagating gravitational degrees  of freedom  with the Compton wavelengths larger than  $L_{N}$. 
  Absence of such new degrees of freedom, would be in contradiction both with breakdown of 
  BH semi-classicality, as well as with impossibility of resolving species identities at distances 
  shorter than $L_N$. Thus, what we are learning is, that introduction of $N$ non-gravitational species alone is impossible.   By consistency, such species must be accompanied by new gravitational species that make sure that gravity goes out of semi-classical regime
  at the scale $L_N$.  Thus, the theory requires an UV-completion at the scale $L_N$. 
   Can such a theory maintain the self-completeness properties of the pure-gravity?
   
  Naively,  from our previous experience such an UV completion looks pretty straightforward.
  We know, that by integrating in new KK gravitons, we can make 
  $L_N$ to be equal to a fundamental Planck length of a higher dimensional theory.  
  Then,  we can expect  that the high-dimensional gravity self-completeness itself,  just in the same way as this happens in a  pure-gravity theory. 
   The problem, however, is the existence of $N$ zero modes, which do not come from the 
   high-dimensional graviton multiplet.  These degrees of freedom require  pre-existence 
   of the ``parent" non-gravitational species in a high-dimensional theory, and the issue of UV-completion is lifted to the next level. 
  
     Let us explain the latter concern by considering an attempt of UV-completing the above 
    scalar example by the BH barrier.   As a first step, we can restore the consistency of the theory with the  BH requirements, by adding $N$ KK gravitons and completing the theory to a $4+d$-dimensional theory with the fundamental Planck length being equal to $L_N$. This step takes care of the consistency with the fact that fundamental length is $L_N$ rather than $L_P$. 
    But the remaining issue is to fit  $N$ zero mode scalars to the completed theory without jeopardizing the property that the strong couplings must be shielded by the BH formation.  
 This is how the new challenge arises.  The most straightforward possibility would be  to promote our $N$ massless scalars into the zero modes of 
 $N$  $4+d$-dimensional fields. But this will not work, since now the fundamental scale 
 of $4+d$ dimensional theory has to be lower than the $4+d$-dimensional Planck length  as 
 \begin{equation}
  L_N^{(4+d)} \, = \, N^{{1 \over d+2}} \, L_{4+d} \, .
  \label{lifted}
  \end{equation}
 So the issue of UV completion procreates. 
  
   We may attempt to introduce species not in form of the high-dimensional fields,  but of zero modes that are localized on some branes.  However, this again does not avoid the problem, 
   since localized species count as much as the bulk ones \cite{giacesar}.  
   Thus,  it is a challenge to consistently introduce non-gravitational species,  
   without creating a strong coupling scale below the Planck length.

 \section{The Role of String Theory}
 
    UV-IR connection exhibited by string theory 
  is strikingly similar  to UV-IR connection of pure Einstein gravity. 
  This  similarity between the deep UV-IR properties of the two theories  suggests some intrinsic connection at the most fundamental level.

 However,  string theory
  introduces a new scale, the string tension scale, $M_s$.   
  Veneziano-type softening of scattering amplitudes starts precisely at the string scale,  which in a weakly coupled string theory can be arbitrarily lower than $M_P$.  
 So,  what is the role of string theory  in UV-completion of gravity?

   In this note, we shall limit ourselves by suggesting a possible line of thought in this direction.  
 One idea  is,  that  in fact string theory and Einstein gravity are non-separable.  In other words,  
a string theory  with order one string coupling $g_s\, \sim \,  1$ is built-in in Einstein gravity.  To put it differently,  by writing down Einstein's action,  we are committing ourselves 
to a string theory. 
 But, where are the string excitations coming from in pure Einstein gravity? 
In this approach, the first string massive excitations have mass $\sim \,  M_P$, and these are precisely quantum states that are suggested by particle-BH transition discussed above. 
 Heavier Regge resonances are simply classical states, and are indistinguishable from classical 
 solutions of Einstein gravity such as heavy BHs or other classical states of IR gravity (e.g.,  loops of long strings).

 Another  idea is,  that string theory is necessary for consistent coupling of Einstein gravity to particle species.  This idea is supported by several findings. First, as we have seen, in the presence of species, a  new scale, $L_N$, parametrically larger than $L_P$ inevitably appears. 
The role of this scale can be naturally played by the string length $l_s$. 
 This matches the previous findings \cite{giacesar},  that string theory is a theory of species with their  
 effective number being $N\, = \, 1/g_s^2$. 


  The latter picture also suggests,  that for $g_s \sim 1$ the difference between string theory and Einstein gravity is essentially erased.  Since, the both descriptions predict existence of  quantum degrees of freedom   around $M_P$.   These may be thought as the species arising as quantum limit of lightest BHs, or equivalently as first string resonances.    
 Only if one needs to introduce many light species, one inevitably has to open up  a finite energy window  below $M_P$, and string theory becomes weakly coupled in order  to accommodate this window. Next, we sketch a possible strategy to figure out how this could happen.
 
 The BH induced UV/IR "bounce" ,
 \eq\label{one1}
 L \rightarrow L'=\frac{L^2_P}{L} \, ,
 \eqx
 lead us to assume that the relevant quantum degrees of freedom controlling the self dual Planckian region are the quantum resonances of mass $M_P$ we get when the BH enters into a quantum regime. In order to make contact with string theory
 we shall make use of existence of  these particle species. 
 At a subplanckian scale $L'$ we will assume that the effective number of species is,
 \eq
 N(L') = \frac{L^2}{L^2_P} \, 
 \eqx
 and therefore on the basis of the results of \cite{giacesar} the effective "string coupling" will be given by
 $g(L') = \frac{1}{\sqrt{N(L')}}$ and the species scale by $L_s =\sqrt{N(L')}L_P$. Note,  that,  as it should be, in this deep UV region we get weak string coupling. If now we consider a state with mass $M=\sqrt{N(L')}M_P$ that is the mass of the effective BH we will create whenever we try to probe scale $L'$ and we write the Planck length in terms of the species scale, we get a string mode with string entropy being the BH entropy. It is amusing to notice, that the transformation from $L_P$ into the species scale $L_s = \sqrt{N(L')}L_P$ is precisely the transformation on the Planck length induced by a string T-duality transformation (\ref{one1}). Indeed if we interpret the UV/IR bounce (\ref{one1}) as a T duality, we will need to change the Planck length as,
 \eq
 L_P \rightarrow L'_P= L_P (\frac{L}{L_P}) \, ,
 \eqx
 which gives  precisely the species relation we have used.
 
 In summary, once we assume that Einstein gravity by itself sets the bound on information storage, a string theory can be naturally built-in by identifying species identities as the information bits.  These bits count  elementary quantum states of mass $M_P$. In this frame of string theory  as of theory of species,  the string coupling flows in the UV to a weak coupling regime.  reflecting the standard way used by string theory  for completing gravity.

A final question that naturally appears is,  if any theory satisfying holography - in the sense of possessing an absolute bound on information storage - should necessarily contain extra quantum states of mass $\frac{1}{L_H}$ with $L_H$ being the holographic scale (the shortest scale 
that can store a minimal information bit) as well as to enjoy some form of duality invariance under the UV/IR bounce $L \rightarrow \frac{L^2_H}{L}$. 


    \section{On Self-UV-Completeness of $11$-Dimensional Supergravity}
    
     An interesting  evidence in favor of  UV-completeness of pure-Einsteinian supergravity
  can be  obtained if  we combine our notion of the BH barrier  with the fact that in a certain consistent decoupling limit string theory is reduced to a pure  gravitational theory, with the 
  Planck length much longer than the string length ($L_s$). 
   We wish to suggest that the  existence of such a decoupling limit,  is an indication that the supergravity theory on its own is UV-complete. 
  
    This is a well-known example  of 
   the strong coupling limit of type $II A$ string theory \cite{Mtheory, matrix}. 
   This example is analyzed in \cite{giacesar1} from the point of view of the behavior of the 
   species scale in strong coupling limit, and the resulting view is, that string theory becomes theory of gravity when other species decouple. 
    
   We shall not repeat all the details of this construction, but only the aspects that are crucial for our  argument.  
   The key feature is the existence of the consistent limit in which string theory decouples and the low energy theory is given by the theory of pure Einstein supergravity in $11$-dimensions.   As explained in \cite{giacesar1},  from the point of view of the remaining  supergravity theory, the species scale in this limit becomes the $11$-dimensional Planck length.     
        
   The decoupling of strings is achieved by taking the limit $g_s \, \rightarrow \, \infty$ and 
   $L_s \, \rightarrow \, 0$, but keeping the $11$-dimensional Planck scale,  
   $M_{11}^{9} \,  \equiv \, M_s^9/g_s^3$,  fixed.    In this limit strings decouple,  but the $D_0$-branes give rise to the  perturbative states. The interpretation of these new perturbative states is, that they  form the KK species 
   of $11$-th dimension that opens up in this limit.   The radius of this $11$-th dimension is, 
\eq\label{one}
R\, = \, g_{s}L_{s} \, .
\eqx
The $10$ and $11$ dimensional Planck lengths satisfy the usual geometric relation,  
\eq
L_{10}^8 =  L_{11}^9/R \, .
\label{massscales}
\eqx
The crucial fact is, that  $L_{11}$ is much larger than the string  length,
\eq\label{three}
L_{11} \, = \, g_{s}^{\frac{1}{3}} L_{s} \,.   
\eqx
  Since at  distances 
larger than $L_{11}$ the theory is a pure-Einstein supergravity,  in the light of our 
BH arguments, $L_{11}$  automatically becomes a shortest length scale of nature. 
 In the other words, any measurement that attempts to retrieve the information at a length scale 
 $L \, \ll \, L_{11}$  bounces us back to the deep IR physics,  corresponding to the 
 length $L_{11}^2/L$,  
 \begin{equation}
 L \, \rightarrow \, {L_{11}^2 \over L} \, .
\label{mtheorybound}
\end{equation}
Now,  since by design, there are no non-gravitational degrees of freedom  until the scale $M_s$, which is infinitely above  $M_{11}$, the  $11$-dimensional supergravity must be UV-complete on its own.  In the other words, by becoming trans-Planckian the stringy physics automatically 
became unreachable, and the remaining gravitational physics has no choice other than completing itself by the available gravitational tools, the BHs.  

   Starting from a consistent UV-complete theory, the ten-dimensional 
   string theory,  we have obtained a pure-gravity theory, with the shortest scale $L_{11}$. 
   As explained above,  because  $M_{11}$ is a boundary between the  quantum 
  and classical objects,  certain quantum degrees of freedom are necessarily "stuck" there.  These 
 quantum degrees of freedom,  together with $11$-dimensional graviton multiplet,  must play the crucial role in self-completing 
  the theory.     
     

   \section{On Entropy Count}
    An independent issue although intimately connected is how to build -with the tools of pure gravity- a microscopic theory that accounts for the BH entropy. The standard answer to this question is that in order to reach this microscopic understanding of BH entropy we need to move into string theory and it is this claim/ hope what normally leads to think of string theory as a way to UV complete gravity. 
 
 It is well known that the string length sets an absolute bound on physical space resolution \cite{GUP2}. This bound as well as the Hagedorn bound on temperatures is intimately related with the extended nature of strings. Moreover this property of strings becomes manifest with the discovery of T-duality. In string theory what prevent us to resolve small scales  ( or to exceed Hagedorn temperature ) is that whenever we pump more energy we end up with a longer string probe. 
As we are pointing out the ancestor of this phenomena is already present in pure Einstenian gravity, where we find that whenever we pump more energy to localize a probe below Planck scale we end up with a BH larger than the Planck length. Thus, independently of any other consideration, quantum gravity should possesses the Planck length as the minimal length.

 In Einstein gravity,  an immediate consequence of this is the existence of a gap between the massless graviton and the first excited state of the theory that should have mass equal to $M_P$. Moreover if we assume holography then the expression of the BH mass spectrum written in "information" terms, namely $M^2 = \frac{N}{L^2_P}$ seems strikingly similar to the standard Regge spectrum of string theory for $L_s=L_P$. 

In spite of this strong similarity we do not need to jump , at least not too quickly, into a string UV completion of gravity where the first excited quantum state appears as a string vibration mode. As already pointed out, we can follow a different path to identify this state using just standard quantum field theory and assuming that Einstenian gravity is complete in the IR and therefore able to fully describe large semiclassical black holes. Indeed we can start with a large BH and identify the first quantum excited state as the remnant of the semiclassical black hole evaporation. 

String theory is avoiding this complicated path involving a classical/ quantum transition and it is replacing it by a sort of RG flow in the string coupling. Indeed we can just take an arbitrary BH and  reach the string state by simply continuously moving the string coupling $g$ appropriately ( keeping $L_s$ fixed ). By this flow in $g_s$ we smoothly ``degravitate" the theory until reaching the point where the string shows up as the Wilsonian UV completion of the theory. In other words, by flowing in $g$ we effectively can move from an effective GR- IR regime (strong coupling in $g_s$) with $L_s << L_P$ into a stringy- UV regime (weak coupling in $g_s$) with 
$L_s >> L_P$. In this sense string theory defines an UV completion in the Wilsonian sense identifying as the UV degrees of freedom the string vibration modes. 

How this string picture can fit with the Einstenian requirement of having $L_P$ as the minimal observable physical length in nature? Part of the answer is contained in the string/BH correspondence \cite{BH/String} or in other words in the microscopic meaning of the BH entropy. In fact if we consider the BH mass written in "information" variables as, 
\eq
M_{BH} = \frac{\sqrt{N}}{L_P} \, ,
\eqx
with the corresponding entropy $S=N$, the understanding of this entropy as string entropy simply requires to define $L_s$ as,
\eq
L_s= \sqrt{N}L_P \, .
\eqx
By doing so the BH mass becomes in "string" variables 
$M_{BH}=\frac{N}{L_s}$, with string entropy $S=N$. It is in the transformation from "information" variables $L_P$ into "string" variables
$L_s$ where we implicitly introduce a string coupling 
$g=\frac{1}{\sqrt{N}}$. However as we have argued in reference \cite{giacesar1},  the previous transformation is simply the definition, within pure Einstenian gravity, of the species scale for a number $N$ of quantum species. In other words, pure Einstenian gravity by setting $L_P$ as the minimal physical length is also setting the bound on information storage and fixing the "information" variables in terms of which the spectrum of BH masses organizes itself in a sort of Regge trajectory. This Regge trajectory becomes stringy once we move into the "species frame", with the species  scale defined relative to a number of species equal to the amount of information. The string flow in $g_s$ that was crucial to the stringy Wilsonian UV completion of gravity becomes a physical flow in information. In pure Einstein gravity this flow in information in encoded in the quantum BH evaporation process.

We can extract some general lessons from the previous picture. In particular, we can imagine as a general feature of any theory that sets by itself the bound on information storage, let us say $L_H$, that mass spectrum could be organized in "information" variables as Regge trajectories. However, the actual decoding of a given amount of information sets the corresponding species scale leading to a string frame with the flow in $g_s$ the flow in information. It seems that Einstenian gravity contains the essential ingredients of this general picture.

   \section{Outlook} 
   
     To summarize, we have argued that the Planck length is the absolute shortest length-scale  of nature, and any attempt of probing the physics beyond it automatically bounces us back to  the physics at large distances.   The maximal physical information $I_{max}(L)$ that can be extracted in any measurement at distance $L \, \ll \, L_P$,  is bounded by the information contained at the horizon of a  classical  BH of radius $L_P^2/L$, 
    \begin{equation} 
      I(L)_{max} \, = \, I(L_P^2/L) \, , 
      \label{info}
  \end{equation}
  and thus, is intrinsically IR in nature. 
  This property indicates that Einstein gravity is self-UV-complete. 
 Of course, the information stored in a classical BH obeys the holographic bound. But this is 
 an intrinsic property  of Einstein BHs rather than an extra assumption.  
 
    Regarding the role of string theory, the emerging conclusion is, that 
 certain strong coupling limit of string theory is built-in in pure Einstein gravity, and that    
 the  role of  weakly-coupled string theory is to consistently couple gravity  to the other particle species, with their number being set by $N \, = \, 1/g_s^2$, and the species scale $L_N$ being set by $L_s$.   String theory decouples together with 
 species, and the resulting limit is a pure gravity theory.  In other words,    
 string theory is the UV-completion of theory of species coupled to gravity above the scale $L_N$ \cite{giacesar1}.

           An interesting question is what is the connection (if any)  of our results with the recent hints of possible perturbative finiteness of $N=8$ supergravity theory\cite{supergravity}.  A priory, since our argument is fully non-perturbative, no perturbative finiteness is necessary.   From our perspective even if loop diagrams are badly divergent,  some resummation that will guarantee the finiteness of the theory  must take place.  At the moment we do not see anything in our arguments that would demand cancellations at loop by loop level.  However,  it may be that the loop-by-loop finiteness is the way the theory  perturbatively  reconciles itself with the existence of the  BH barrier.          
                  
                  
    \subsection{Non-Wilsonian ``Higgless" UV-Completion by Gravity?}                           
                                   
     Finally, an interesting phenomenological question is, whether one can use gravity 
   for the UV completion of the Standard Model without any need of a Higgs particle?
     For this,  first one needs to add to Einstein gravity the extra gravitational degrees of freedom in order  to  stretch the fundamental Planck length $L_*$ all the way to the electroweak distances.  In such a case the 
 strong coupling in scattering of longitudinal $W$-bosons will be UV-completed by classical  BHs. 
   This can certainly be achieved by introducing large extra dimensions,  as it was already done for solving the hierarchy problem\cite{add}.  However,  the question is not just lowering the scale $M_*$, but whether one can get away without introducing the Higgs.  This idea would be in spirit somewhat similar to the ``Higgless" models \cite{higgsless}, except   the known Higgless 
   models rely on  Wilsonian completion of the theory, in which  the Higgs particle is substituted by other quantum  degrees of freedom that restore unitarity up to sufficiently high scales. 
   
   In contrast,  our idea is to look for non-Wilsonian UV-completion in which  in deep-UV the quantum degrees of freedom become classical states.   
  To fix terminology,  completion that we are trying to suggest  will not be truly Higgless, in the 
  sense that the Higgs-like  degree of freedom  will inevitably appear  as the state at the boundary 
 that separates classical BHs from  quantum degrees of freedom.  
 To be more precise, once we couple the standard model without the Higgs to gravity, the Higgs particle will appear automatically at the scale $M_*$, as the latest stage of the evaporation of a BH with the 
 quantum numbers of the Higgs doublet.   Indeed, such a BH can always be formed by 
 scattering quarks and leptons with appropriate gauge quantum numbers. The existence of a particle-like state in the spectrum will result from the latest stages of the evaporation, at which 
 the BH is indistinguishable from a heavy particle.   
 So in this sense, gravity automatically provides
 a composite Higgs in form of a quantum BH, even if initially there was no Higgs scalar in the spectrum of the SM species.   In this respect this idea also shares some similarity with the idea of the top quark condensate \cite{top}. 
 
 Of course,  there are many phenomenological questions that make it unclear if such a scenario can  ever work,  the electroweak precision parameters being an immediate concern.  Another question is, why is the BH-type Higgs condensing?  
  We shall postpone answering these question for future.

  \vspace{5mm}
\centerline{\bf Acknowledgments}

The work of G.D. was supported in part by Humboldt Foundation under Alexander von Humboldt Professorship,  by European Commission  under 
the ERC advanced grant 226371,  by  David and Lucile  Packard Foundation Fellowship for  Science and Engineering and  by the NSF grant PHY-0758032. 
The work of C.G. was supported in part by Grants: FPA 2009-07908, CPAN (CSD2007-00042) and HEPHACOS P-ESP00346.
 The results of this paper were presented at the "Fundamentals of Gravity" workshop at LMU (Munich).   We would like to thank organizers for providing such an opportunity.

\end{document}